\documentclass[12pt]{article}
\usepackage{graphicx}
\usepackage{cite}
\usepackage{color}
\newcommand{\balpha}{\mbox{\boldmath $\alpha $}}

\newcommand{\bsigma}{\mbox{\boldmath $\sigma $}}
\begin{document}
\begin{titlepage}
\title{{\bf The Harari-Shupe Observation without Preons - a Glimpse of Physics to Come?}}
\author{
{Piotr \.Zenczykowski \footnote{email: piotr.zenczykowski@ifj.edu.pl}}\\
{\it Division of Theoretical Physics}\\
{\it The Henryk Niewodnicza\'nski
Institute of Nuclear Physics}\\
{\it Polish Academy of Sciences}\\
{\it PL 31-342 Krak\'ow, Poland}
}
\maketitle
\begin{abstract}
We argue that one has to distinguish between the Harari-Shupe model (HSM) 
and the Harari-Shupe observation (HSO). The former --- in which quarks and leptons are viewed as composite objects built from confined fermi\-on\-ic subparticles (`rishons') --- is known to be beset with many difficulties. The latter may be roughly defined as this part of the HSM that really works. We recall that the phase-space Clifford-algebra approach leads to the HSO without any of the HSM problems and  discuss in some detail how this is achieved.  The light which the phase-space-based view sheds on the HSO sets then a new direction along which the connection between space and particle properties could be studied and offers a glimpse into weird physics that probably lurks much deeper than the field-theoretical approach of the Standard Model. 

\end{abstract}
\vfill

\end{titlepage}

\section{Introduction}
It was a cold morning some hundred thousand years ago.
Naoh~\footnote{The intelligent Neanderthal 
                 featured in "The Quest for Fire", who contributed to the gene pool of modern humans.}
 accidentally hit a rock with his flint axe and small stone flakes flew off.
He looked at the edge of his axe with fear and, angry at his clumsiness, struck one of these flakes, mindlessly
staring at even smaller pieces as they appeared. Suddenly, a thought crossed his mind.
He struck again one of the just produced flakes with his precious axe and watched the new emerging chips. Then, he repeated the procedure. 
His theory worked! There were flakes within flakes within flakes... forever.
 
At present, driven by the desire to seek deeper and deeper into the structure of matter, we keep building better and harder axes to split such tiny chips into even tinier ones. 
And yet, at the same time we feel intuitively that such divisibility cannot go on indefinitely.
It cannot be `turtles all the way down'.
We accept the Democritean resolution of this Kantian antinomy: at some point we must reach chips 
which are not divisible any longer, true `atoms' moving in the continuous background space.
Yet, our evolutionary background and the macroscopic everyday experience sit in us deeper than the Democritean tenet. They induce us to continually shift down the level at which divisibility stops. This is how we went from stone flakes to molecules to chemical elements to hadrons to quarks.
Although the nature of consecutive chips is altered at each level crossing along this chain, such a change constitutes a mild modification only of the intuitive picture deeply ingrained in our minds.~\footnote{Strictly speaking, the transition from hadrons to quarks goes somewhat beyond what this picture allows.
The reason is that quarks, as we have good reasons to believe, are forever confined, and only their
conglomerates -- the hadrons -- satisfy the intuitive definition of separate chips.}
With each such successful step down, our confidence in the continuing divisibility receives another boost.
As a result, when at a new level we are confronted with 
the existence of several similar objects that may be grouped into Mendeleev-type tables,
many followers of Democritus retreat into the position of Naoh.

Today, when we have reached the level of the Standard Model (SM), an analogous Mendeleev-type table 
is constructed for leptons and quarks, the fundamental particles of the SM. Our earlier successes 
suggest then that the  level of Democritean indivisibility be shifted another step down.
This is the conceptual basis of the attempts to build quarks and leptons out of a novel brand of
subparticles: the preons \cite{PatiSalam}. The most famous of such attempts
is the Harari-Shupe rishon model (HSM) \cite{HarariShupe}. The model is very economic
and so appealing in its internal symmetry that it is hard not to believe that it contains
an element of truth. Yet, at the same time, the model - and indeed the whole idea of preons - is beset 
with numerous difficulties which, in the eyes of disbelievers, strongly suggest that preons do not exist. 

Can it be that both believers and disbelievers are at least partially right? We think so. We believe that while the original
formulation of the HSM does not provide an adequate description of nature, the model contains a very important element of truth (to be defined later). Furthermore, a simple assumption should exist which would
produce this element only, while simultaneously avoiding the  difficulties of the HSM. 

Such an approach was proposed  in a series of publications 
\cite{ZenAPPB1,ZenPLB,ZenJPA,ZenITJP,ZenBook} over the last several years. The 
basic underlying idea consists in the
replacement of the non-relativistic (NR) arena 
of 3D space and time with the 6D phase space. 
This phase-space approach seems to be often regarded
as a scheme that provides a theoretical justification for the preon models and, 
given current lack of belief in such models, it is ignored. 
Such an attitude is based on a complete misunderstanding. Let us state it clearly:
{\it the phase-space-based scheme is not intended as a basis for any version of the preon model}. It is just to the contrary:
{\it the phase-space scheme shows how one can get the relevant element of truth (which we call the Harari-Shupe observation, or HSO for short)
without any subparticle structure of quarks and leptons}.

In this paper we attempt to clarify these points, trying to be as simple as possible.
Thus, we avoid discussing those details of the phase-space scheme that could unnecessarily complicate our presentation and refer to the original papers for the more involved calculations and further arguments. 
We will first describe the HSM together with its difficulties and define what we mean by  the HSO. Then, we present the main elements 
of the phase-space scheme showing how it reproduces the HSO. Subsequently, we move on to analyse point by point all the main
difficulties of the HSM that result from the supposed preon substructure of leptons and quarks, 
and explain why these difficulties do not appear in the phase-space scheme. Finally, we present the 
 weird spatial picture that the phase-space scheme suggests for the connection between quark and hadron levels of the description of matter and argue that it provides a glimpse 
into how space and time are related to some underlying pregeometric structure.

\section{The Approaches}
\subsection{The Harari-Shupe Model}
The original Harari-Shupe model assumes that there are only two types of truly fundamental spin-1/2 particles: the `rishon' $T$ of charge  $Q_T=+1/3$ (in units of proton charge), and the rishon $V$ of charge $Q_V=0$. Naturally, these particles are accompanied by their antiparticles:
$\bar{T}$ and $\bar{V}$. A composite particle may have half-integer spin when
the number of its constituent spin-1/2 subparticles is odd (thus at least three). Accordingly, in the HSM the ordinary spin-1/2 elementary particles 
(leptons and quarks) are built out of three confined rishons.  Specifically, $\nu_e$, $u_R$, $u_G$, $u_B$, $e^+$, $\bar{d}_R$, $\bar{d}_G$, $\bar{d}_B$, the eight elementary particles of a single SM generation, are identified with ordered rishon combinations shown
in Table \ref{HSM}. The corresponding  antiparticles, i.e.  $\bar{\nu}_e$, $\bar{u}_R$, 
$\bar{u}_G$, $\bar{u}_B$,  $e^-$, $d_R$, $d_G$, $d_B$, are built as in Table \ref{HSM} with $V$ replaced by $\bar{V}$ and $T$ by $\bar{T}$. The HSM is very economic and exhibits some additional nice features. For example, since there is an equal number of protons and electrons in the Universe, and since a proton-electron pair contains an equal number of rishons and antirishons (4$T$, 4$\bar{T}$, 2$V$ and 2$\bar{V}$), it follows that
nature is fully symmetric under matter-antimatter interchange at the rishon level.

\begin{table}[h]
\caption{The Harari-Shupe Model: rishon structure of the $I_3=+1/2$ members of a single SM generation}
\begin{center}
{\begin{tabular}{cccccccc} \hline
 $\nu_e $&$u_R$&$u_G$&$u_B$
&$e^+ $&$\bar{d}_R$&$\bar{d}_G$&$\bar{d}_B$\rule{0mm}{6mm}\\
 $\displaystyle VVV$&$\displaystyle TTV$&
$\displaystyle TVT $&$\displaystyle VTT$&$TTT$&$VVT$&$VTV$&$TVV\vphantom{\frac{1}{j_k}}$\rule{0mm}{6mm}\\
\hline
\end{tabular}}
\end{center}
\label{HSM}
\end{table}

\subsection{Criticisms of HSM}
The main problems with the original HSM are:
\begin{itemize}
\item With rishons being spin-1/2 particles one expects that their spins could also be added so as to form spin-3/2 partners of leptons and quarks. However, such states are not observed.  
\item There is a serious problem with rishon statistics: identification of colored quarks with the ordered (i.e. not antisymmetrized) combinations of rishons is in conflict with the assumed fermionic nature of rishons.
\item There is a problem with the origin of lepton/quark masses. In the currently dominant field-theoretical picture, rishons are imagined as particles confined to distances smaller than the maximum acceptable quark or lepton size of about $10^{-16}$ cm. Consequently, momentum  uncertainties of such subparticles should be huge, and their energies --- much larger than electron or light quark masses. Why, therefore, are those masses so small?
\item There is no naturally appearing $SU(3)$ color symmetry: the relevant three-rishon combinations are identified with colored quarks solely on account of the triplicity of states built of two different rishons.
\item No underlying rishon-binding dynamics is proposed. 
Specifically, it is not explained why\\ 
   \phantom{XX}1) $TTT$, $VVV$ are free, but $TVV$, $TTV$ are confined,\\
   \phantom{XX}2) $TT\bar{T}$, $VV\bar{V}$ are not observed (even as confined objects),\\
   \phantom{XX}3) the observed free particles are built from $TTT$, $VVV$, $\bar{T}\bar{T}\bar{T}$, $\bar{V}\bar{V}\bar{V}$, $T\bar{T}$ and $V\bar{V}$ only,
\item the model admits unobserved baryon-number-violating processes such as
$u+u \to e^+ + \bar{d}$ which is possible via exchange of rishons, e.g.
     $TT{\bf{V}}+TV{\bf{T}} \to TT{\bf{T}} + TV{\bf{V}}$.\\
\end{itemize}
These problems have been addressed in various papers formulated within the general subparticle paradigm. Their main idea was to endow the approach with some additional complex structure that removes the shortcomings of the original model.  

\subsection{The Harari-Shupe Observation}
The essence of the Harari-Shupe model consists in the
observation that the charges of the eight fundamental fermions of $I_3=+1/2$ can be constructed in a specific additive way from only two charges $Q_T=+1/3$ and $Q_V=0$, as shown
in Table \ref{HSO}. The word `specific' means that the three distinguishable orders of adding 
$Q_T$, $Q_T$, and $Q_V$ (i.e. $Q_T+Q_T+Q_V$, $Q_T+Q_V+Q_T$, $Q_V+Q_T+Q_T$) are indeed treated as such (and likewise for $Q_T \leftrightarrow Q_V$).

\begin{table}[h]
\caption{The Harari-Shupe observation: additive structure of the charges for the $I_3=+1/2$ members of a single SM generation}
\begin{center}
{\begin{tabular}{cccc} \hline
 $\nu_e $&$u_R$&$u_G$&$u_B$\rule{0mm}{6mm}\\
 $ 0+0+0$&$ \frac{1}{3}+\frac{1}{3}+0$&
$ \frac{1}{3}+0+\frac{1}{3} $&$ 0+\frac{1}{3}+\frac{1}{3}$ $\vphantom{\frac{1}{\frac{f}{g}}}$\rule{0mm}{6mm}\\
\hline
$e^+ $&$\bar{d}_R$&$\bar{d}_G$&$\bar{d}_B$\rule{0mm}{6mm}\\
$\frac{1}{3}+\frac{1}{3}+\frac{1}{3}$&$0+0+\frac{1}{3}$&$0+\frac{1}{3}+0$&$\frac{1}{3}+0+0$$\vphantom{\frac{1}{\frac{f}{g}}}$\rule{0mm}{6mm}\\
\hline
\end{tabular}}
\end{center}
\label{HSO}
\end{table}

In order to explain the HSO, Harari and Shupe assume the subparticle paradigm. Within that paradigm the component charges necessarily reside on subparticles. It should be clear, however, that this assumption of the existence of subparticle components of quarks and leptons constitutes a superfluous addition to the observation made in Table \ref{HSO}. Although the charge of an elementary particle is built in Table \ref{HSO} via the addition of some `components', this does not imply that these components reside on 
separate particles.~\footnote{Compare the case of a stick: although the number of its ends is naturally obtained as a sum $1+1=2$, a component of this sum (`1') does not sit {\it alone} on any individual `substick'.}
 There may exist a principle different from the subparticle paradigm, which leads to Table \ref{HSO} without the associated baggage of the HSM problems.

The general philosophy of extended rishon-like models is to supplement the HSM with an additional drawback-correcting structure. In other words, one first introduces preons together with all the drawbacks they induce and then adds the correction mechanism.
{\it Wouldn't it be simpler and more in accord with Occam's razor to avoid both the introduction of problems in the first place and the subsequent introduction of the mechanism of their avoidance?} As we shall see, this is what the Clifford algebra phase-space approach does actually achieve.

\subsection{Clifford algebra phase-space approach}
The Clifford algebra phase-space approach is based on a generalization of Dirac's trick. In the simplest nonrelativistic case this trick consists in the linearization of 
momentum vector square ${\bf p}^2=p_1^2+p_2^2+p_3^2$. Namely, one writes ${\bf p}^2$ as a product of two identical factors linear in vector ${\bf p}$:
\begin{equation}
\label{NRlinearize}
{\bf p}^2 = ({\bf p}\cdot {\bf A} ) ({\bf p}\cdot {\bf A}) 
\end{equation} 
where ${\bf A} =(A_1,A_2,A_3)$ represents some momentum-independent vector-like object. The absence of terms proportional to $p_mp_n$ (with $m \ne n$) on the l.h.s. of 
Eq. (\ref{NRlinearize}) requires that $A_k$ satisfy anticommutation rules 
\begin{equation}
\label{anticom}
A_mA_n+A_nA_m=\delta_{mn},
\end{equation} 
which may be reproduced if one takes $A_k = \sigma_k$, where $\sigma_k$ are  Pauli matrices.
With the spin operator of spin-1/2 particles being given in terms of Pauli matrices as
${\bf s}=\frac{1}{2}\bsigma$ (we choose units such that $\hbar = 1$), the requirement that
\begin{equation} 
\label{pA}
{\bf p} \cdot {\bf A}
\end{equation} 
be invariant under ordinary rotations links then the quantum concept of spin with the rotational properties of vectors in macroscopic 3D space. 

One can provide various philosophical and symmetry-based arguments (see e.g. Ref. \cite{ZenBook}) that Eq. (\ref{pA}) should be generalized to its phase-space extension:
\begin{equation}
\label{NR2lin}
{\bf p} \cdot {\bf A}+{\bf x} \cdot {\bf B},
\end{equation}
where ${\bf x}$ denotes position vector, while ${\bf A}$ and ${\bf B}$ are two dimensionless vector-like objects independent of momentum and position. 
In order to make the above addition of momentum and position terms possible, proposal (\ref{NR2lin}) requires the introduction of a new constant of nature $\kappa$ (just as with $\hbar$, we set it equal to 1 by an appropriate choice of units), of dimension [momentum / length] so that ${\bf x}$ may be measured in momentum units.  Together with the quantum constant
$\hbar$ of dimension [momentum $\times$ length], the two constants set the absolute scale of both momenta and distances.~\footnote{This does not necessarily mean that $\hbar/\kappa$ defines a universal minimal quantum of distance.} The absolute scale of masses is defined when the speed of light $c$ is added.

Elements  $A_m$ and $B_n$
satisfy a straightforward generalization of anticommutation relations (\ref{anticom}), i.e. 
\begin{eqnarray}
A_mA_n+A_nA_m=B_mB_n+B_nB_m&=&\delta_{mn},\nonumber\\
A_mB_n+B_nA_m=0, \label{NRCliff}
\end{eqnarray}
and may be represented by $8 \times 8$ matrices, whose explicit form may be chosen as 
\begin{eqnarray}
A_m&=&\sigma_m\otimes\sigma_0\otimes\sigma_1,\nonumber\\
B_n&=&\sigma_0\otimes\sigma_n\otimes\sigma_2.
\end{eqnarray}
In addition to the mutually anticommuting $A_m$ and $B_n$, 
the algebra composed of (1) a unit element, (2) $A_k$ and $B_l$, and (3) all antisymmetric multiple products of $A_m$ and $B_n$ (i.e. the Clifford algebra in question) contains one additional element which anticommutes with all $A_m$ and $B_n$. It is constructed from $A_m$ and $B_n$ as 
\begin{equation}
B=iA_1A_2A_3B_1B_2B_3=\sigma_0\otimes\sigma_0\otimes\sigma_3.
\end{equation}

The phase-space analog of Eq. (\ref{NRlinearize}) is obtained by squaring expression (\ref{NR2lin}) under the quantum condition that $[x_m,p_n]=i\delta_{mn}$.
One finds:
\begin{equation}
\label{px2}
({\bf p} \cdot {\bf A}+{\bf x} \cdot {\bf B})({\bf p} \cdot {\bf A}+{\bf x} \cdot {\bf B})={\bf p}^2+{\bf x}^2+R,
\end{equation}
where 
\begin{equation}
R=-\frac{i}{2}\sum_k[A_k,B_k]=\sum_k\sigma_k\otimes\sigma_k\otimes\sigma_3.
\end{equation}
The appearance of nonzero $R$ is due to the fact that position $x_m$ and momentum $p_n$ do not commute for $m=n$.

The fundamental conjecture of the phase-space approach  consists in the
identification of the charge operator $Q$ with an appropriately modified (scaled) product (\ref{px2}):
\begin{equation}
\label{chargeQ}
Q=\frac{1}{6}\left[({\bf p}^2+{\bf x}^2)_{vac}+R\right]B,
\end{equation}
where $({\bf p}^2+{\bf x}^2)_{vac}=3$ is the lowest (vacuum) eigenvalue of 
${\bf p}^2+{\bf x}^2$.
Thus, formulas (\ref{px2}, \ref{chargeQ})  propose a link between the properties of phase space and the concept of quantized charge. In other words, just as the properties of quantized spin  are tied to rotations in ordinary 3D space, so the properties of quantized charge  are conjectured to be tied to certain transformations in 6D phase-space.
As we will show, it is assumption (\ref{chargeQ}) that replaces the subparticle paradigm of the HSM.
Eq. (\ref{chargeQ}) may be rewritten as
\begin{equation}
Q=I_3+\frac{Y}{2},
\end{equation}
with the third component of (weak) isospin $I_3$ and (weak) hypercharge $Y$ defined as:
\begin{eqnarray}
I_3&=&\frac{B}{2}=\frac{1}{2}\,\sigma_0\otimes\sigma_0\otimes\sigma_3,\nonumber\\
Y&=&\frac{1}{3}RB=\frac{1}{3}\sum_k\sigma_k\otimes\sigma_k\otimes\sigma_0\equiv Y_1+Y_2+Y_3. 
\end{eqnarray}
On the r.h.s. above we have introduced three `partial hypercharges' ($k=1,2,3$):
\begin{equation}
\label{partialhypercharge}
Y_k\equiv -\frac{i}{6}[A_k,B_k]B = \frac{1}{3}\sigma_k\otimes\sigma_k\otimes\sigma_0.
\end{equation}
In the phase-space language the ordinary three-dimensional rotations and reflections are naturally
understood as simultaneous operations on vectors ${\bf p}$ and ${\bf x}$ 
(and their matrix counterparts ${\bf A}$ and ${\bf B}$). It is easy to check that operators $I_3$ and $Y$ are invariant under these operations.~\footnote{Indeed, $R$ is proportional to the difference of two scalar products:
${\bf A}\cdot{\bf B}$ and  ${\bf B}\cdot{\bf A}$, while $B$ is proportional to a product of two mixed (pseudoscalar) products: $A_1A_2A_3$ and $B_1B_2B_3$.} 
In addition, one finds that 
\begin{equation}
[Y_k,Y_m]=[Y,Y_m]=[I_3,Y_m]=[I_3,Y]=0.
\end{equation}
Thus, the eigenvalues of all $Y_m$, $Y$ and $I_3$ may be simultaneously specified.
 One gets 
\cite{ZenAPPB1,ZenPLB,ZenJPA,ZenITJP,ZenBook} 
\begin{eqnarray}
I_3&=&\pm \frac{1}{2},\nonumber\\
Y_k&=&\pm \frac{1}{3}.
\end{eqnarray}
Yet, the eigenvalues of $Y_1$, $Y_2$, $Y_3$ are not independent of one another. One finds the constraint $Y_1Y_2Y_3=-1/27$ and a restricted set of eigenvalues of $Y$: 
\begin{eqnarray}
Y&=& -1, +\frac{1}{3}, +\frac{1}{3}, +\frac{1}{3}.
\end{eqnarray}
For the antiparticles one has to substitute $Y_k \to -Y_k$. The allowed combinations lead to eight possibilities for $\{Y_1,Y_2,Y_3\}$,  which are gathered in Table \ref{Ys}. 
\begin{table}[h]
\caption{Alternative version of the Harari-Shupe observation: the allowed decompositions of the eigenvalues of $Y/2$ into the eigenvalues of
$Y_1/2$, $Y_2/2$, and $Y_3/2$. Upper and lower rows correspond to particles and antiparticles respectively (both labelled with the names of the eight $I_3=+1/2$ members of a single SM generation).}
\begin{center}
{\begin{tabular}{cccc} \hline
$\nu_e $&$u_R$&$u_G$&$u_B$\rule{0mm}{6mm}\\
  $ -\frac{1}{6}-\frac{1}{6}-\frac{1}{6}$&$ +\frac{1}{6}+\frac{1}{6}-\frac{1}{6}$&
$ +\frac{1}{6}-\frac{1}{6}+\frac{1}{6} $&$ -\frac{1}{6}+\frac{1}{6}+\frac{1}{6}$ $\vphantom{\frac{1}{\frac{f}{g}}}$\rule{0mm}{6mm}\\
\hline
$e^+ $&$\bar{d}_R$&$\bar{d}_G$&$\bar{d}_B$\rule{0mm}{6mm}\\
$+\frac{1}{6}+\frac{1}{6}+\frac{1}{6}$&$-\frac{1}{6}-\frac{1}{6}+\frac{1}{6}$&$-\frac{1}{6}+\frac{1}{6}-\frac{1}{6}$&$+\frac{1}{6}-\frac{1}{6}-\frac{1}{6}$$\vphantom{\frac{1}{\frac{f}{g}}}$\rule{0mm}{6mm}\\
\hline
\end{tabular}}
\end{center}
\label{Ys}
\end{table}

One notes strict correspondence between Table \ref{Ys} and the original HSO:
Table \ref{HSO} is obtained from Table \ref{Ys} simply by adding $\Delta=+1/6$ to each eigenvalue of $Y_k/2$. Thus,  the phase-space approach indicates 
that the original HSO could be equally well formulated in terms of
the eigenvalues of $Y$ and $Y_k$. The correspondence between the original (Table \ref{HSO}) and the new (Table \ref{Ys}) version of HSO is ($k=1,2,3$ labels the position of rishon in the HSM state):
\begin{eqnarray}
Q_V=0&\leftrightarrow & Y_k=-\frac{1}{3},\\
Q_T=+\frac{1}{3}&\leftrightarrow & Y_k=+\frac{1}{3}.
\end{eqnarray}
In the HSM the antiparticles of the set given in Table \ref{HSM} (i.e. the eight fermions of $I_3=-1/2$) are composed of antirishons $\bar{T}$ and $\bar{V}$.  The corresponding versions of Tables \ref{HSO} and \ref{Ys} are obtained by simply changing all of the signs in their entries. Therefore, the connection between the two antirishon versions of the HSO is obtained by adding $\Delta=-1/6$ to each eigenvalue of $Y_k/2$.

\section{Disappearance of HSM difficulties}
Let us now discuss the HSM difficulties and their absence in the
phase-space scheme.

\subsubsection*{Absence of spin-3/2 partners of leptons and quarks}
Assumption (\ref{chargeQ}) which connects the concept  of quantized
charge with the symmetry properties of nonrelativistic phase-space avoids the use of fermionic subparticles. Thus, one cannot infer the existence of spin-3/2 partners of ordinary leptons and quarks. However, even if we tried to augment the phase-space scheme
with underlying spin-1/2 rishons \`a la HSM, we would necessarily fail for such an introduction of spin-1/2 subparticles is impossible. Indeed, within the phase-space scheme, a partial hypercharge $Y_k$ {\it cannot} be assigned to a spin-1/2 subparticle. It cannot be done for the simple reason that $Y_k$ refers to {\it one} (the $k-$th) direction in ordinary 3D space only, while any discussion of spin requires the inclusion of all three spatial directions.

\subsubsection*{Absence of antisymmetrization}
Assumption (\ref{chargeQ}) does not introduce fermionic subparticles. Consequently, the `composite states' are simply not there, and there is no way to attempt their antisymmetrisation. There is also no reason for any antisymmetrization: the phase-space scheme gives the HSO directly, i.e. no modifications of the scheme are needed.
\subsubsection*{The problem of mass}
The absence of subparticles naturally prevents the appearance of any paradox that might result from the confrontation  of the smallness of lepton/quark masses with the huge momentum uncertainty of confined constituent rishons. 

Yet, while the original HSM mass paradox disappears, the problem of mass acquires an altogether different and very interesting look. Indeed, with the presence of two phase-space constants $\kappa$ and $\hbar$, the absolute scale of masses is fixed when the speed of light $c$ is added. Thus, one may expect that the phase-space scheme, when properly developed, should contitute a basis for a totally different approach to the problem of mass \cite{Born}.
In fact, this problem constituted one of the main reasons behind the development of the scheme. This reason may be analysed starting from the HSO in the modified form of Table \ref{Ys}. Namely, one observes that the permutation
\begin{eqnarray}
\label{nutouR}
\nu_e \leftrightarrow u_R,&u_G \leftrightarrow u_G, & u_B \leftrightarrow u_B,
\end{eqnarray}
may be achieved by the following interchange among $Y_k$'s:
\begin{eqnarray}
Y_1 \to Y'_1=-Y_2, & Y_2\to Y'_2=-Y_1, & Y_3 \to Y'_3=Y_3.
\end{eqnarray}
This interchange may in turn be obtained from Eq. (\ref{partialhypercharge}) via the following operation on elements $A_m$ and $B_n$ (among other possibilities):
\begin{eqnarray}
A_1\to A'_1=B_2,~~~~~ & A_2 \to A'_2=-B_1, ~~~~~&  A_3 \to A'_3=A_3,\nonumber \\
B_1\to B'_1=A_2,~~~~~ & B_2 \to B'_2=-A_1, ~~~~~&  B_3 \to B'_3=B_3.
\end{eqnarray}
The above operation has its natural counterpart in phase space:
\begin{eqnarray} 
p_1\to x_2,~~~~~ & p_2 \to -x_1, ~~~~~&  p_3 \to p_3,\nonumber \\
x_1\to p_2,~~~~~ & x_2 \to -p_1, ~~~~~&  x_3 \to x_3,
\label{rotphsp}
\end{eqnarray}
which constitutes a specific joint rotation (by $\pi/2$) in phase-space planes $(p_1,x_2)$ and $(x_1,p_2)$. Eq. (\ref{rotphsp})
interchanges some momenta coordinates with some position coordinates. Thus, via Eq. (\ref{rotphsp}),
a symmetry transformation (\ref{nutouR}) that changes a lepton into a quark, transforms the standard Dirac Hamiltonian of a lepton, i.e. $\balpha \cdot {\bf p}+\beta m$, into a translationally noninvariant expression. Consequently, in the phase-space scheme a colored quark is not described by the Dirac Hamiltonian. Since this  contradicts the basic assumption of the Standard Model,  it might be argued that the whole phase-space idea should be immediately discarded. 
It turns out, however, that the above drawback may be turned into a virtue. In fact,
the issue just raised may be considered a hint on how  the concept of quark mass and the relation that it is supposed to fulfill are to be reinterpreted. A detailed analysis of how the concept of quark mass was originally introduced into the Standard Model shows that such a reinterpretation is possible.
For more details see Refs \cite{ZenBook,ZenDICE2014}. Below, we will use only some of the hints that Eq. (\ref{rotphsp}) does suggest.  
\subsubsection*{Absence of SU(3)}
As the basic equations (\ref{px2}, \ref{chargeQ}) show, the phase-space scheme naturally introduces the group of rotations in the 6D phase space  (i.e. the $SO(6)$ group). From mathematics we know that this group contains $U(1)\otimes SU(3)$ as a subgroup. Simple analysis shows then that  the $SU(3)$ transformations do not affect the lepton sector  in Table \ref{Ys} (i.e. a lepton is a singlet under $SU(3)$ transformations), but transform between themselves the sectors of three colored quarks in this Table (see e.g. Ref. \cite{ZenPLB}). Thus, contrary to the HSM case, in the phase-space scheme there is a color $SU(3)$ group that transforms between $TTV$, $TVT$, and $VTT$ sectors of the original HSO. A more elaborated connection to quantum chromodynamics is missing, however.
\subsubsection*{Binding of rishons}
In the original HSM no rishon dynamics is proposed. The resulting rishon-binding problem manifests itself
in various ways in different multirishon states. Its general resolution in the phase-space scheme is based on three observations:
\begin{enumerate} 
\item that $T$ and $\bar{V}$ both correspond to $Y=+1/3$, while $V$ and $\bar{T}$ to $Y=-1/3$, 
\item that the distinction between $T$ and $\bar{V}$ (or between $V$ and $\bar{T}$) consists in shifts by $\Delta=+1/6$ for the charges of rishons $T$, $V$ and by $\Delta=-1/6$ for the charges of antirishons $\bar{T}$, $\bar{V}$,  
\item that there is a specific connection between phase-space variables and lepton/quark sectors. 
\end{enumerate}
We proceed now to the discussion of various ways in which the binding problem reveals itself.
\begin{itemize}
\item {\it $TTT$, $VVV$ are free, but $TVV$, $TTV$ are confined}\\
The difference between these two types of three-rishon states is twofold.
First, one observes that the total hypercharge $Y$ of these states is integer ($\pm 1$) for
$TTT$, $VVV$ states and fractional ($\pm1/3$) for $TVV$, $TTV$. As a resolution of the problem one could therefore simply postulate that states with fractional values of $Y$ are individually unobservable. However, since confinement refers to the macroscopic classical behavior of particles, its deeper discussion must involve the classical concept of particle separation in ordinary 3D space. 
Consequently, and this is the second point, the difference in the spatial behavior of particles with  integer and fractional values of hypercharge should be correlated with the phase-space variables which appear in the relevant (lepton and quark) Hamiltonians . From Eq. (\ref{rotphsp}) we see that for quarks the ordinary momentum in the lepton Hamiltonian is replaced by a more general form of `canonical momentum' in which some momentum coordinates are replaced by the corresponding position coordinates. Since the appearance of these position coordinates leads to translationally noninvariant Hamiltonians for individual quarks, the requirement of translational invariance at our classical macroscopic level (satisfied for the $TTT$ and $VVV$ combinations by assuming the Dirac form of lepton Hamiltonians) naturally forbids the appearance of $TVV$, $TTV$ combinations as free particles.  

\item {\it $TT\bar{T}$, $VV\bar{V}$ are not observed}\\
First, for these combinations their total values of $Y$ are fractional 
which brings us close to the cases of $TTV$ and $VVT$.
Second (and more importantly), these combinations correspond to the total values of $I_3$ equal to $\Delta(T)+\Delta(T)+\Delta(\bar{T})=
\Delta(V)+\Delta(V)+\Delta(\bar{V})=1/6+1/6-1/6 =1/6$ 
which is unacceptable as $I_3$ has to be integer or half-integer.
Thus, in the phase-space scheme the $TT\bar{T}$ and $VV\bar{V}$ combinations simply do not exist.

\item {\it observed free particles are built from $TTT$, $\bar{T}\bar{T}\bar{T}$, $VVV$, 
$\bar{V}\bar{V}\bar{V}$, $T\bar{T}$, and $V\bar{V}$ only}\\
Indeed, the individually separable particles are \\ 
\phantom{xxx} 1) four leptons, built from rishons as $TTT$, $\bar{T}\bar{T}\bar{T}$, $VVV$ or $\bar{V}\bar{V}\bar{V}$
and\\ 
\phantom{xxx} 2) mesons and baryons, built from rishons as shown below.\\
Specifically, mesons are composed of rishons as
\begin{eqnarray}
I_3=0,~~~~~~~u\bar{u}&=&(TTV,\bar{T}\bar{T}\bar{V})\to (T\bar{T})^2(V\bar{V}),\nonumber\\
I_3=1,~~~~~~~u\bar{d}&=&(TTV,VVT)\to (TTT)(VVV),\nonumber\\
I_3=0,~~~~~~~d\bar{d}&=&(\bar{V}\bar{V}\bar{T},VVT)\to (T\bar{T})(V\bar{V})^2,\nonumber\\
\label{mesons}
I_3=-1,~~~~~d\bar{u}&=&(\bar{V}\bar{V}\bar{T},\bar{T}\bar{T}\bar{V})\to(\bar{T}\bar{T}\bar{T})(\bar{V}\bar{V}\bar{V}),
\end{eqnarray} 
which, in the second version of HSO, are all of the same form (of total $Y=0$): 
\begin{equation}
(Y_1+Y_2+Y_3,Y_1+Y_2+Y_3)=(1/3+1/3-1/3,-1/3-1/3+1/3).
\end{equation}
Similarly, baryons are composed of rishons as  
\begin{eqnarray}
I_3=3/2~~~~~~~uuu &=&(TTV,TTV,TTV)\to (TTT)^2(VVV),\nonumber\\
I_3=1/2~~~~~~~uud &=&(TTV,TTV,\bar{V}\bar{V}\bar{T})\to(TTT)(T\bar{T})(V\bar{V})^2,\nonumber\\
I_3=-1/2~~~~~udd&=&(TTV,\bar{V}\bar{V}\bar{T},\bar{V}\bar{V}\bar{T})\to (T\bar{T})^2(V\bar{V})(\bar{V}\bar{V}\bar{V}),\nonumber\\
\label{baryons}
I_3=-3/2~~~~~ddd&=&(\bar{V}\bar{V}\bar{T},\bar{V}\bar{V}\bar{T},\bar{V}\bar{V}\bar{T})\to
(\bar{T}\bar{T}\bar{T})(\bar{V}\bar{V}\bar{V})^2,
\end{eqnarray}
which, in the second version of HSO, are all of the same form (of total $Y=1$):
\begin{eqnarray}
&(Y_1+Y_2+Y_3,Y_1+Y_2+Y_3,Y_1+Y_2+Y_3)&\nonumber\\
&=(1/3+1/3-1/3,1/3+1/3-1/3,1/3+1/3-1/3).&
\end{eqnarray}
When compared with leptons and states composed of leptons (e.g. $e^+\nu_e$=$(TTT)$ $(VVV)$), which all are free states, the rishon composition of hadrons differs by the possible presence of $(T\bar{T})$ and $(V\bar{V})$ factor states.
 Since in the phase-space scheme we associated $T$ ($\bar{T}$) with $\Delta=+1/6~ (-1/6)$ contribution to $I_3$, one gets integer (zero) value of total $I_3$  for
the $(T\bar{T})$ state, but neither integer nor half-integer values for
the $(T)$ and $(TT)$. Analogous results hold for 
states $(V\bar{V})$, $(V)$, $(VV)$ as well as for $(V\bar{T})$ etc.
In addition, the $(T)$, $(TT)$, and other similar states have fractional values of $Y$, while for $(T\bar{T})$ one has $Y=0$.
Thus, the absence of $(T)$, $(TT)$ and other similar factor states on the right hand side of Eqs (\ref{mesons}, \ref{baryons}) constitutes merely a translation of the conditions that 
for the observed free particles the values of their $Y$ are integer and those of $I_3$ are integer or half-integer.

Obviously, as the concept of particle freedom refers to particle's behavior in space, one still needs to connect the above quantum number argument with a space picture.  In fact,
we have already pointed out that in the phase-space scheme, contrary to the case of leptons,  the Hamiltonians of individual quarks violate translational invariance. This was interpreted as equivalent to quarks not being free particles. It is therefore of great interest to see whether   certain  conglomerates of quarks can be made to appear free in the phase-space scheme, i.e. if (and --- if yes ---  how) the translational invariance  could be restored for hadronic states. We will discuss this issue in Section \ref{glimpse}.
\end{itemize}
\subsubsection*{Baryon number violation}
The HSM rishons are introduced as ordinary (even though confined) particles. Thus,  the states composed of rishons may exchange their components upon sufficiently close contact without any obvious penalty.  
As a result, the state $TTV+TVT$ may rearrange its rishons into $TT{\bf T} + TV{\bf V}$ , i.e. transition $u+u' \to e^+ + \bar{d}$ appears possible. Unfortunately for the HSM,
this is a baryon-number-violating process which has not been observed in nature.
     
In the phase-space scheme such a transition requires an `exchange' of partial hypercharges (e.g. $Y^u_3 \leftrightarrow Y^{u'}_3$):
\begin{eqnarray}
&(Y^u_1=\frac{1}{3},Y^u_2=\frac{1}{3},Y^u_3=-\frac{1}{3}) +
 (Y^{u'}_1=\frac{1}{3},Y^{u'}_2=-\frac{1}{3},Y^{u'}_3=\frac{1}{3})& \nonumber \\ &\to& \nonumber \\
&(Y^e_1=Y^u_1,Y^e_2=Y^u_2,Y^e_3= Y^{u'}_3)
+ (Y^{\bar{d}}_1=Y^{u'}_1 ,Y^{\bar{d}}_2=Y^{u'}_2,Y^{\bar{d}}_3=Y^{u}_3).&
\end{eqnarray}
Such an interchange assigns to individual partial hypercharges a particle-like independence that is {\it not} built into the phase-space scheme. The partial hypercharges that define a given particle cannot be traded between different particles.
Thus, baryon-number violation does not occur in the phase-space scheme.

\section{A glimpse of physics to come?}
\label{glimpse}

As already stressed, integer (fractional) values of $Y$ are associated with free (confined) particles.
In order to proceed with the discussion of this connection between the value of hypercharge and the spatial behavior of elementary particles,  including the emergence and behavior of hadrons as envisaged in the phase-space approach, we introduce now two reasonable assumptions:
\begin{enumerate} 
\item a connection between particle-antiparticle conjugation and phase-space variables (to treat the case of mesons which involve both quarks and antiquarks) and 
\item a plausible prescription for how to combine the canonical momenta of quarks.
\end{enumerate}
For simplicity, we will restrict our discussion to hadronic states composed of quarks and antiquarks of a single flavor (e.g. $u$ and $\bar{u}$) and the association of their canonical momenta with phase-space variables. Inclusion of $d$ and $\bar{d}$ is discussed in Ref. \cite{ZenITJP}.\\

The connection between particle-antiparticle conjugation and phase-space variables may be established by a simple analysis of the assumed invariant behavior of position-momentum commutation relations under the operations of $P$ and $T$. We have:
\begin{eqnarray}
P: ~~~i \to +i,~~ & {\bf p} \to {\bf p'}=-{\bf p},~~~~ & {\bf x} \to {\bf x'}=-{\bf x},\nonumber\\
T: ~~~i \to -i,~~ & {\bf p} \to {\bf p'}=-{\bf p},~~~~ & {\bf x} \to {\bf x'}=+{\bf x}.
\end{eqnarray}
If $CPT$ is an identity operation, then $C$ is {\it represented in the phase-space language}~\footnote{It is a fallacy to identify an abstract description of reality (such as e.g. the field-theoretical approach to elementary particles, however successful it is) with this reality itself. Consequently, the representation of $C$ may depend on the language chosen.}
by a product of $P$ and $T$, i.e. by
\begin{eqnarray}
C: ~~~i \to -i,~~ & {\bf p} \to {\bf p'}=+{\bf p},~~~~ & {\bf x} \to {\bf x'}=-{\bf x}.
\label{Cconj}
\end{eqnarray}
Consequently, in order to go from particles to antiparticles we have to 1) change everywhere the sign in front of the imaginary unit $i$ (which leads to complex conjugate representations and the opposite sign of charge), and 2) reverse the signs with which position coordinates enter into the relevant formulas (which leads to an unorthodox interpretation of antiparticles in macroscopic terms, a phase-space-based counterpart of the Feynman-St\"uckelberg interpretation).\\

In order to discuss in some detail the issue of spatial properties of states composed of quarks we turn first to states composed of ordinary, freely observable particles, such as e.g. a two-lepton state. As a whole, this state is characterized by its total momentum, i.e. by the {\it sum} of the momenta of its constituents. Similarly, when two hadrons collide and a resonance is formed, its momentum is taken to be the sum of the momenta of initial hadrons.
This additivity of momenta, an established and elementary property of any system of free particles or macroscopic objects, is so deeply ingrained in our minds that in the standard approaches we take it for granted that it applies also to quarks. In the phase-space scheme, however, 
quarks are described with the help of `canonical momenta' in which some momentum components are replaced by position components. Should we apply then the additivity principle to the ordinary momenta or to the canonical momenta?
Our fundamental conjecture is that it is more natural to combine the canonical momenta.\\

We stress that, contrary to the case of free particles, the additivity of the physical momenta of individual colored quarks in a given hadron {\it cannot be} {\it confirmed in a strictly experimental way}. Indeed, due to confinement, a colored quark cannot be observed (as an individual free particle) and, consequently, its physical momentum cannot be measured. As a result, in current approaches to strong interactions this momentum is  merely assigned to an individual quark. This is done with the help of both theory and phenomenology. There is no rigorous theoretical transition between the level of measured hadron momenta and the level of quark momenta. Consequently, there is also no rigorous way to check the additivity of quark physical momenta, a property which is again merely assigned to any system of quarks. We suspect that the phenomenological problems encountered in the standard  description of various properties of hadrons in terms of their quark structure (in space or in momentum space) appear precisely because of the conflict between these standard assignments and our conjecture that one should combine quarks' canonical momenta. \\

\subsection{Translational invariance and confinement}

If we accept that one should apply the additivity principle to canonical momenta, several interesting conclusions follow. First, in agreement with Eqs (\ref{rotphsp},\ref{Cconj}) we note that $u_R$ and $\bar{u}_R$ are associated with canonical momenta as follows (the bar signs over phase space variables for the antiquark distinguish them from those for the quark):
\begin{eqnarray}
u_R & \leftrightarrow & (-x^R_1,+x^R_2,p^R_3), \nonumber\\
\label{RRbarcanmomenta}
\bar{u}_R & \leftrightarrow & (+\bar{x}^R_1,-\bar{x}^R_2,\bar{p}^R_3).
\end{eqnarray} 
The additivity principle -- when applied to canonical momenta -- leads then to the total canonical momentum of the $u_R\bar{u}_R$ system being 
\begin{equation}
\label{RRbar}
(+\bar{x}^R_1-x^R_1,x^R_2-\bar{x}^R_2,p^R_3+\bar{p}^R_3),
\end{equation}
 which is a translationally invariant expression. 
Thus, if confinement is identified with  the lack of translational invariance, the quark-antiquark system is not confined.
On the other hand, for the $u_{R1}u_{R2}$ system the additivity principle leads to the total canonical momentum being
$(-x^{R1}_1-x^{R2}_1,x^{R1}_2+x^{R2}_2,p^{R1}_3+p^{R2}_3)$, which is still a translationally non-invariant expression. 

If we replace one of the two $u_R$ quarks with a quark of another color, e.g. $u_G$, we should combine the r.h.s of the first line of Eq. (\ref{RRbarcanmomenta}) and the r.h.s. of the corresponding expression for a green quark, i.e.
\begin{eqnarray}
u_G & \leftrightarrow & (+x^G_1,p^G_2,-x^G_3),
\end{eqnarray}
where the signs are determined by cyclicity. A natural way to combine the two canonical momenta seems to be
\begin{eqnarray}
u_Ru_G & \leftrightarrow & (+x^G_1-x^R_1,+x^R_2,-x^G_3,p^G_2,p^R_3),
\end{eqnarray}
which expression is still translationally noninvariant.
We conclude that translational invariance cannot be restored by forming a system
of two quarks. Thus a diquark is necessarily confined. 
Since by cyclicity we have for the blue quark and antiquark:  
\begin{eqnarray}
u_B & \leftrightarrow & ( p^B_1,-x^B_2,+x^B_3),\nonumber\\
\bar{u}_B & \leftrightarrow & ( \bar{p}^B_1,+\bar{x}^B_2,-\bar{x}^B_3),
\end{eqnarray}
we note that for the $u_Ru_G$ system the position components in the second and third direction, i.e. $x^R_2$ and $x^G_3$, enter with same signs (respectively $(+,-)$ ) as for the blue antiquark ($\bar{x}^B_2$, $\bar{x}^B_3$) and with the opposite signs as for the blue quark. Thus, as far as the translationally non-invariant components are concerned, the $u_Ru_G$ system behaves like the $\bar{u}_B$ antiquark. \\

Under translations, therefore, the $u_Ru_Gu_B$ system should behave just like the $\bar{u}_Bu_B$ system, for which a translationally invariant expression similar to (\ref{RRbar}) may be written.
In fact, for the $u_Ru_Gu_B$ system, our combination prescription suggests the form
\begin{equation}
\label{triangle}
(p^B_1,p^G_2,p^R_3,x^G_1-x^R_1,x^R_2-x^B_2,x^B_3-x^G_3),
\end{equation}
which is explicitly translationally invariant. Hence, apart from the nonconfined $u\bar{u}$
quark-antiquark states (mesons) there should exist nonconfined three-quark $uuu$ states (baryons).
 Thus, the proposed prescription for combining the canonical momenta of quarks leads to the conclusion similar to that following from the standard group-theoretical argument according to which only $SU(3)_{color}$ singlet states ($q\bar{q}$, $qqq$, ... )  are  observable as free separable particles, while $q$, $qq$, and other color nonsinglet states are confined. 
Our conjecture on how to combine the canonical momenta may be therefore viewed as corroborated by quark confinement and the existence of unconfined mesons and baryons. Although the above arguments may be regarded as highly simplistic, we believe that they reflect the appearance of confinement fairly well and 
provide a glimpse of conditions that a future better formulation of such ideas should fulfill.

While the phase-space scheme naturally leads to the appearance of the $SU(3)$ color group, which is definitely superior to the case of the HSM, the standard QCD gauge structure would still have to be imposed ad hoc. On the other hand, if we interpret position differences $x^G_1-x^R_1$ etc. as describing components of interquark strings, the phase-space scheme naturally connects the internal color structure of hadrons with their string-like properties, a result that in QCD requires the solution of the confinement problem. 
 
\subsection{Rotational covariance and the concept of a point}
The expressions suggested  by the prescription for combining the canonical momenta of quarks exhibit peculiar features as far as their rotational properties are concerned. Indeed, the total canonical momentum of the $u_R\bar{u}_R$ state is not rotationally invariant. In order to construct a  rotationally covariant description of meson momenta one has to consider both $u_R\bar{u}_R$ as well as $u_G\bar{u}_G$ and $u_B\bar{u}_B$ states. While one may easily write the formal  expression $(p^B_1+\bar{p}^B_1,p^G_2+\bar{p}^G_2,p^R_3+\bar{p}^R_3)$, its rotational covariance clearly requires quarks of different colors to conspire.
Since in the baryon case the total baryon momentum suggested by the combination prescription is described by $(p^B_1,p^G_2,p^R_3)$, a somewhat similar conspiration between the  $u_R,u_G,u_B$ quarks is needed also for baryons.

Note that we are talking here about the rotational properties of the {macroscopic} concepts of positions and momenta for quarks and the systems of quarks, not about the properties of the quantum concept of spin for quarks and the system of quarks (we accept that quark spins are adequately described by the standard Pauli matrices). 
The weird nature of spatial conspiracy that is here suggested to exist between quarks of a given hadron should not discourage us. After all, the proposed prescription for combining the canonical momenta of quarks leads to a novel view on confinement.
 Thus, the encountered oddity 
should not be taken as an argument against the proposed scheme. To the contrary, 
we think that the required spatial conspiracy of quarks does shed light on the very nature of macroscopic space, a question that lies totally outside of the standard field-theoretical approach.

 The spatial arguments of the fields (i.e. positions or momenta)
are not of a microscopic nature \cite{Wigner}. They
 provide a classical macroscopic reference frame for the quantum particles. 
Thus, as argued by David Finkelstein \cite{Finkelstein},
 the field-theoretical description of elementary particles is a hybrid one: it involves both the macroscopic classical continuous variables (positions or momenta), and the strictly quantum variables such as spins and other discrete quantum numbers.
In line with the idea of emergent spacetime, particle positions (and momenta) are supposed to constitute the  concepts that appear only when a vast number of strictly quantum systems interact \cite{Zimmerman}. Recall now that hadronic positions or momenta can be measured but those of individual colored quarks cannot be. Consequently,  one may argue that
the whole standard and familiar structure of the 3D space (together with its rotational and Minkowskian aspects~\footnote{A way to introduce special relativity is discussed in Ref. \cite{ZenDICE2014}.}) becomes operationally sufficiently well defined only at the colorless (hadronic) level. As a result, an adequate description of the quark-level structure may deviate from the currently dominant macroscopically-driven geometrical ideas and require some kind of `pregeometry' \cite{Wheeler}. We think that
the condition of rotational covariance of a quark-level description of macroscopic hadronic variables
provides us with a  glimpse on how such pregeometry might look like.

All this does not mean that  quarks  cannot be associated and described at some higher level of the phase-space scheme with the standard bispinor fields $q(x)$ defined on the ordinary spacetime manifold, as it is customarily done in the Standard Model. Indeed, the {\it colorless} aspects of quark behavior 
(e.g. couplings of color-singlet quark currents to photons or weak gauge bosons)
have to be describable in terms of such macroscopically-covariant fields.
However, as in the phase-space scheme
the Hamiltonian of an individual colored quark appears to be a translationally noninvariant object, the relevant bispinor  quark fields cannot satisfy standard Dirac on-mass-shell equations. Therefore, the orthodox approach that uses Dirac quarks must constitute an approximation.

We stress that  the introduction of the concept of position as an argument of quark field $q(x)$ finds its experimental justification via the hadronic level observables only. The translation of hadron-level observations to the standard quark-level picture requires additional assumptions. 
In the QCD description of deep inelastic scattering the necessary translation  is achieved with the help of the phenomenological interface of `structure functions'.  
With $x$ being satisfactorily defined at hadronic and higher levels only, such additional assumptions which extrapolate the concept of point to the `interior' of hadrons may be unjustified. In particular,
imagining quarks as located at specific points of an underlying 3D background space and confined to a region of this space may (and - in our opinion - should) be regarded as a simplifying and approximate description of nature.~\footnote{ In fact, there are strong indications from the phenomenology of baryon spectroscopy \cite{CapstickRoberts} (see also Ref. \cite{ZenDICE2014}) that the standard quark model / lattice QCD description of excited baryonic states is an idealization that misses a crucial aspect of their internal spatial structure.}

\section{Concluding remarks}
\subsubsection*{No preons}
In this paper we argued in some detail that the Harari-Shupe observation should and may be explained without the introduction of preons. We also pointed out that the phase-space approach provides precisely such an explanation.
The continuing attempts to subdivide elementary particles again and again (and to treat space as infinitely divisible) do show the strength of our hundred thousands years old evolutionary inheritance, but are against the spirit of the proper resolution of the problem, as started by Democritus and elaborated later by Heisenberg \cite{HeisenbergPhysToday} (see below).
The existence of a philosophically sound, very economic, and successful explanation of the HSO in terms of phase-space symmetries strongly suggests that the level of quarks and leptons constitutes the lowest level of the divisibility of matter.

\subsubsection*{The changing meaning of the word `to divide'}
A somewhat deeper reflection regarding the concept of division consists in the realisation 
 that the Democritean idea of indivisible atoms consists in choosing the most crucial step --- in the long chain of conceptual changes concerning the notion of divisibility --- as the only such change.
 Indeed,  with each subsequent step down --- when going along the complexity ladder from the macroworld to the world of elementary particle --- the word `division' is stripped of some of the macroscopic attributes we usually associate with it (this is an extension of the original idea discussed by Heisenberg in Ref. \cite{HeisenbergPhysToday}).  For example, during the transition from the molecular to the atomic level the chemical properties largely dissappear. Yet, the property of separability in space is not modified for a long series of such consecutive steps. This changes only when the transition from the hadronic to the quark level is effected. At this stage, the macroscopic concept of divisibility loses its crucial feature: the quark `flakes' are no longer macroscopically separable in space.~\footnote{At this point the change in the concept of divisibility is so big that one may repeat after  Heisenberg: ``The word `dividing' loses its meaning.'' \cite{HeisenbergPhysToday}.}
As a result, the standard vision supplied by our imagination, i.e. that of individual quarks being separated by ordinary space, need not be wholly correct: it may constitute an over-simplified idealization. Obviously, there are many serious indications that hadrons are composed of quarks. Neither this view nor the successes of the relevant field-theoretical description are challenged here. Yet, at the same time there are various hints that the precise nature of hadron compositeneness still evades our understanding. Accordingly we think that the current views on the mechanism of quark confinement should be regarded as an insufficient approximation to reality.

The word `to divide' can be used at the lower levels of the divisibility ladder provided we keep it being appropriately redefined at each consecutive step down. The redefinition required at the hadron/quark transition is already substantial, but - as the Standard Model demonstrates - one can accept a theory that circumvents the loss of macroscopic spatial separability.
On the other hand, the step to the rishon level seems to require such a drastic redefinition of the word `to divide' that it is not appropriate to use that word any longer: in particular, in the phase-space scheme it is not just the concept of separability that is lost --- what seems to evaporate is the very concept of the underlying space. In other words, space appears to be a concept
that emerges from some pregeometric `rishon' level.
 
\subsubsection*{The scale of emergence}
As is well known, the idea that spacetime is an emergent macroscopic concept constitutes a starting point in contemporary approaches to quantum gravity. This general idea of spacetime emergence is likely to be correct. After all, we know of many other examples of various properties which emerge when one climbs the ladder of complexity. There seems to be no reason why other concepts, spacetime included, should not conform to this general rule.

What is unorthodox in our view is the distance scale at which, as we argue,  the effects of spacetime emergence can be seen. In the  approaches to quantum gravity, spacetime is thought to emerge at the diminuitive Planck length scale, some 20 orders of magnitude below the hadronic length scale.
If one accepts that the HSO brings out an element of truth, then the standard problems with its preon-based explanation suggest that an important step in that emergence occurs at the rishon-to-quark/lepton transition.
Yet, standard no-preon arguments do not really specify the distance scale relevant for spacetime emergence. 
On the other hand, the phase-space scheme explanation of the HSO does go further. The conjectured additive treatment of position coordinates and its association with the idea of confinement essentially
suggest that important aspects of spacetime emergence are completed only with the next step up the complexity ladder --- i.e. with the quark-to-hadron transition. Although this is a distance scale that is much larger 
than the Planck scale, the idea cannot be regarded as disproved by the successes of the current Standard Model description of elementary particles, which constitutes a {\it field-theoretical idealization} and approximation to reality, and must not be identified with nature \cite{HeisenbergIdealization}. Specifically, the idea advocated here is that (1) the concept of spacetime point, an undisputed input into all field-theoretical formalisms, is an emergent concept, and that (2) one can learn more about this emergence by a deeper understanding of the quark/hadron transition. Note that we do not claim that spacetime emerges just at the hadronic distance scale. As the alocal nature of quantum correlations suggests \cite{Bell,Norsen2}, we suspect that it emerges at {\it all} distance scales.
We think, however, that important hints as to the mechanism of this emergence could be unravelled by a deeper understanding of the quark-to-hadron transition.
\\

\vfill

\vfill
\end{document}